\documentstyle[aps,pre,multicol,epsfig,amssymb]{revtex} 

\def\u{{\mbox{\boldmath$u$}}}

\def\k{{\mbox{\boldmath$k$}}}

\def\sk{{\mbox{$\hat s$}}}

\def\la{{\langle}}
\def\ra{{\rangle}}
\def\om{{\omega}}

\newcommand{\bm}[1]{\mbox{\boldmath $#1$}}
\newcommand{\be}{\begin{equation}}
\newcommand{\ee}{\end{equation}}
\newcommand{\bea}{\begin{eqnarray}}
\newcommand{\eea}{\end{eqnarray}}

\begin{document}

\title{Inverse velocity statistics in two dimensional
turbulence}

\author{ L. Biferale$^{(a,d)}$, M. Cencini $^{(b,e)}$, A. Lanotte$^{(c,d)}$,
  D. Vergni $^{(b)}$}
\address{$(a)$ Dipartimento di Fisica, Universit\`a di Roma ``Tor Vergata'',\\
Via della Ricerca Scientifica 1, I-00133 Roma, Italy}
\address{$(b)$ Dipartimento di Fisica, Universit\`a di Roma ``La Sapienza''
 and INFM, \\
P.le Aldo Moro 2, I-00185 Roma, Italy}
\address{$(c)$ CNR, ISAC - Sezione di Lecce, Str. Prov. Lecce-Monteroni Km 1.200, I-73100 Lecce, Italy}
\address{$(d)$ INFM, Unit\`a di Tor Vergata, Via della Ricerca Scientifica 1, I-00133 Roma, Italy}
\address{$(e)$ CNRS, Observatoire de la C\^ote d'Azur, B.P. 4229, F-06304 Nice cedex 4, France}

\maketitle

\begin{abstract}
 We present a numerical study of two-dimensional turbulent flows in
 the enstrophy cascade regime, with different large-scale forcings and
 energy sinks. In particular, we study the statistics of
 more-than-differentiable velocity fluctuations by means of two
 recently introduced sets of statistical estimators, namely {\it
 inverse statistics} and {\it second order differences}.  We show that
 the $2D$ turbulent velocity field, $\bm u$, cannot be simply
 characterized by its spectrum behavior, $E(k) \propto
 k^{-\alpha}$. There exists  a whole set of 
 exponents associated to the non-trivial smooth fluctuations of the
 velocity field at all scales.  We also present a numerical
 investigation of the temporal properties of $\bm u$ measured in
 different spatial locations.
 \\ PACS: 47.27.Eq, 47.27.Gs \hfill\break
\end{abstract}

\begin{multicols}{2}
\section{Introduction}
\label{sec:intro}
Many natural phenomena display complex fluctuations over a wide range
of spatial and temporal scales. Complexity usually manifests in the
non-Gaussian properties of probability distribution functions
(PDF). When PDFs at different scales do not collapse by a simple
rescaling procedure one speaks about intermittency \cite{Frisch}. Such
non-trivial rescaling properties may be exhibited by PDFs' tails or
peaks, or both \cite{CLMV01}.  When intermittency manifests in the
PDF's tails, it means that regions of very intense bursting activity
are present. This is typical of three dimensional turbulent flows,
where the velocity field is strongly intermittent and rough
\cite{Frisch}.\\ However, there are examples of other important
natural phenomena which develop simple PDF's tails but non-trivial
PDF's cores.  PDF's peaks are associated to laminar fluctuations,
i.e., ``smooth'' variations of the field.  A physically relevant
example is offered by two dimensional turbulent flows where the
presence of long living coherent structures, e.g., vortices, is very
well known (see Figure~1). Two dimensional turbulence is characterized
by two different transport processes\,: an inverse energy cascade from
the forcing scale to larger scales and a direct enstrophy cascade from
the forcing scale to smaller scales \cite{kraichnan_2d}. Inverse
energy cascade shows a non intermittent Kolmogorov 1941 scaling for
the velocity field \cite{SY93,PT98,BCV00}. On the contrary, in the
direct cascade non-trivial vorticity fluctuations have been observed
in dependence on the large scales characteristics of the flow
\cite{SBL89,ott,ohkitani,boffy}. In addition, velocity fluctuations 
in the direct enstrophy cascade regime are particularly interesting for 
geophysical and astrophysical sciences \cite{lesieur}. In this regime, the
velocity field is differentiable, therefore the standard analysis
(customarily applied in 3d turbulence), based on moments of velocity
increments (the so-called structure functions) is poorly
informative. Indeed structure functions are dominated by the
differential component of the signal:
\begin{equation}
S_p(r)= \la [s(x+r)-s(x)]^p \ra \sim r^{p}\,,
\label{eq:uno}
\end{equation}
where with $s$ we indicate either the $u_x$ or the $u_y$ velocity
fields component. It is worth stressing that the scaling behavior
(\ref{eq:uno}) does not imply that the velocity statistics is
trivial. For example, it is well known that in the enstrophy cascade
regime the energy spectrum shows a power law $E(k) \propto
k^{-\alpha}$ with $\alpha \ge 3$, which is the signature of
significant more-than-differentiable velocity fluctuations. Hence,
subdominat contributions to the $s(x+r)-s(x) \propto r$ behavior must
be present and, in principle, detectable. The triviality of the
scaling (\ref{eq:uno}) it is just the consequence of not having chosen
the suitable observable. Therefore, to extract interesting information
on the statistics of smooth signals, new statistical tools are
needed.\\ 
Recent contributions have shown that laminar events are
optimally characterized in terms of their exit-distance statistics,
also known as {\it inverse statistics}
\cite{Je99,BCVV99,entropyphysicad,entropyphysicad2,prl_2d}.  In a
nutshell, in such approach one ``inverts'' the usual way of looking at
signals. Standard analysis studies the statistics of
signal increments over a certain spatial (or temporal) interval; the
exit-distance approach looks at the statistics of spatial (temporal)
intervals necessary to observe a given signal increment.  Another possibility
to study smooth signals is to eliminate the differentiable
contribution by looking at signal {\it Second Differences} (SD), i.e.,
$(s(x+r)-2s(x)+s(x-r))$ as suggested in \cite{Eyink}.\\ In this paper
we extend a previous exploratory investigation \cite{prl_2d} of the
inverse statistics of velocity fields in the enstrophy cascade regime
of $2D$ turbulence, and we compare it with results obtained by Second
Difference statistics on the same flows.  We present both exact
analytical results for the exit-distance probability density functions
of 1D Gaussian signal, and a set of numerical investigations of spatial and
temporal statistics of 2D turbulent flows. The main result is the
identification of highly non-trivial contributions to the
more-than-differentiable velocity fluctuations.  We also introduce a
set of exponents which characterizes smooth behaviors beyond that of
the energy spectrum, $\alpha$, $E(k) \propto k^{-\alpha}$.  \\ The
paper is organized as follows. In section~\ref{sec:1}, we recall some
known results on 2D turbulent flows in the presence of a drag mechanism at
large scales. In section~\ref{sec:2}, we introduce the main 
observables, i.e., the inverse structure functions and the the Second
Difference structure functions\,: we first apply them to the analysis of
stochastic signals with a given spectrum $E(k)\sim k^{-\alpha}$, for
which we are able to establish some exact results. Then in
section~\ref{sec:3}, we present the spatial statistics of laminar
fluctuations of the two dimensional velocity field $\bm u$ obtained by
direct numerical simulations (DNS). In section~\ref{sec:4}, we perform a
temporal analysis of the velocity field on fixed spatial
locations. Section~\ref{sec:5} is devoted to conclusive remarks.
\begin{figure}
\label{fig:snap}
\epsfxsize=2.9truecm
\epsfysize=1.5truecm
\centerline{\epsfig{figure=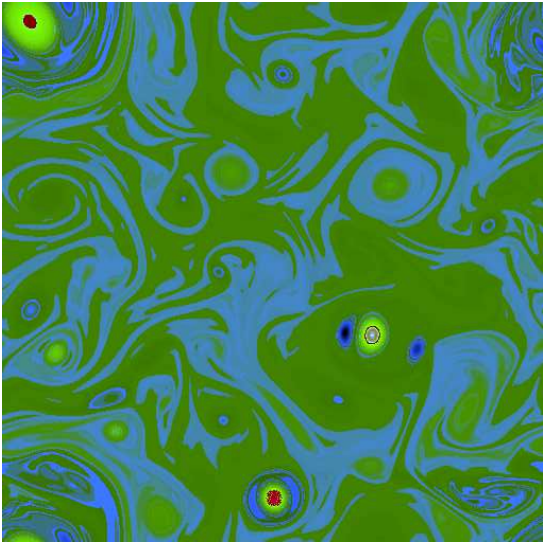,width=0.40\textwidth,height=0.4\textwidth,
angle=90}}
\vspace{0.2cm}
\caption{A snapshot of the vorticity field. Colors are coded according
to the intensity of the vorticity field from the minima of $\om$ (black) to the maxima (clear yellow,white). DNS
have been performed by a standard dealiased pseudo-spectral algorithm, 
over a double periodic square domain of size $L=2\pi$, at resolution
$512^2$ and $1024^2$.  As customary enstrophy is dissipated at small
scales with an hyper-viscosity of order $4$, while energy is removed
at large scales, to avoid piling up on the smallest mode, using
different drags (see table). We considered a Gaussian, white-in-time
large-scale forcing restricted on wave numbers, $4 < |k_f| \leq 6$.  }
\end{figure}
\section{Two dimensional turbulence}
\label{sec:1}
\noindent 
As far as the inertial range of scales for the enstrophy cascade of
two-dimensional turbulence is concerned, previous experimental,
theoretical and numerical studies have shown that the statistics is
strongly influenced by large-scale phenomena. Indeed {\it more than
smooth} spectra $E(k) \sim k^{-\alpha}$ with $\alpha>3$, depending on
the characteristics of the forcing and of the large-scale dissipation,
have been reported~\cite{SBL89,ott,ohkitani}.  Recently, new results
have clarified the problem in the special case of the large scale
energy sink given by a linear (Eckman) friction \cite{ott,boffy}.  
 We recall that the presence of an energy sink at large
scales is conceptually justified by the necessity of avoiding the pile
up of energy on the gravest mode as a result of the inverse energy
cascade \cite{SY93} and it is physically motivated in terms of the
friction to which a fluid is subjected in the Eckman layer
\cite{Eckman1,RW00}. \\ The strong influence of large-scale phenomena
in the whole enstrophy cascade range is believed to be a consequence
of non local interactions (in Fourier space). Another property
associated to the enstrophy cascade is the velocity field
smoothness. The aim of this paper is to discuss a new set of
observable suitable to highlight the statistics of all those
fluctuations which appear as a sub-leading contribution to the smooth
differentiable behavior, $u(x+r)-u(x) \sim r$. \\ Let us now briefly
fix the notation. In terms of the scalar vorticity $\omega={\bf
\nabla} \times {\bm u}$, the equation of motion can be written as
\begin{equation}
\label{eq:NS}
\partial_t \omega + \bm u \cdot {\bf \nabla} \omega =\nu_q \Delta^q \omega -
\beta_{\rho} \Delta^{-\rho}\omega + F,
\end{equation}
where $\nu_q$ and $\beta_{\rho}$ ($q,\rho \ge 0$) are the coefficients
of the generalized dissipations, namely the hyper-dissipative and the
hypo-friction terms respectively. The former removes 
enstrophy at small scales and the latter removes energy at large scales.  $F$ is the vorticity source term acting at large scales. In
Fig.~1, we show a typical snapshot of the vorticity field obtained by
direct numerical simulation of Eq.~(\ref{eq:NS}). As one can see the
vorticity field is characterized by filamental structure over a wide
range of scales.\\ According to the classical prediction
\cite{kraichnan_2d}, the velocity field should exhibit a
Batchelor-Kraichnan spectrum, $E(k)\sim k^{-3}(\ln(k))^{-1/3}$. The
dimensional estimate is observed in a bunch of numerical and
experimental measurements \cite{PJT99,Bo93}.  However, in the
literature there are reported numerous situations
\cite{SBL89,ott,ohkitani,boffy} where different velocity spectra have
been measured\,: $E(k)\sim k^{-\alpha}\,$, with the exponent $\alpha$
larger than $3$ and dependent on the forcing and drag mechanisms. In
the case of linear friction, ($\rho=0$), it is known that vorticity
statistics is intermittent. In such a case, it has been recently
clarified \cite{ott,boffy} that, at scales small enough, vorticity
behaves as a passive scalar. In addition, the dependency of the
spectrum slope on the linear-friction coefficient has been understood
\cite{ott,boffy}. Except for the situation with a large-scale linear
friction, there is no general theory for the scaling properties of
$2D$ turbulent flows in the presence of different large-scale drag
mechanisms (see also \cite{TS02}).  \\ Let us
therefore present the way we analyzed 2D turbulent flows with general
large-scale physics and the related results on their statistics.
\section{Inverse and direct statistics for smooth signals}
\label{sec:2}
In this section we introduce the inverse statistics and the second
difference structure functions. We start applying them to the analysis
of stochastic one-dimensional signals with a given spectrum $E(k)\sim
k^{-\alpha}$.  For the sake of simplicity we limit our
discussion to signals with $3\leq \alpha < 5$, for which we are able to
establish some exact results. To be more precise, we consider smooth
random signals built as follows
\begin{equation}
   s(x) = \sum_k \sk(k) e^{i(kx +\theta_k)}\,,
   \label{smooth}
\end{equation}
where $|\sk(k)|^2 \sim k^{-\alpha}$ and $\theta_k$ are random phases,
uniformly distributed in $[0,2\pi)$.  When $3 \le \alpha < 5$ the
signal is smooth but only one time differentiable. Hence, moments of
its differences over any increment $r$ always possess a differentiable
scaling (\ref{eq:uno}), while moments of order $p\le -1$ do not exist.
\subsection{Inverse statistics}
\label{sec:2.1}
For a generic smooth one-dimensional signal $s(x)$, looking at inverse
statistics consists in measuring moments of the distance, $r(\delta s)$, 
necessary to observe in the signal a double exit (forward and backward)
through a barrier $\delta s$. 
\begin{figure}
\label{fig:method}
\epsfxsize=2.9truecm
\epsfysize=1.5truecm
\centerline{\epsfig{figure=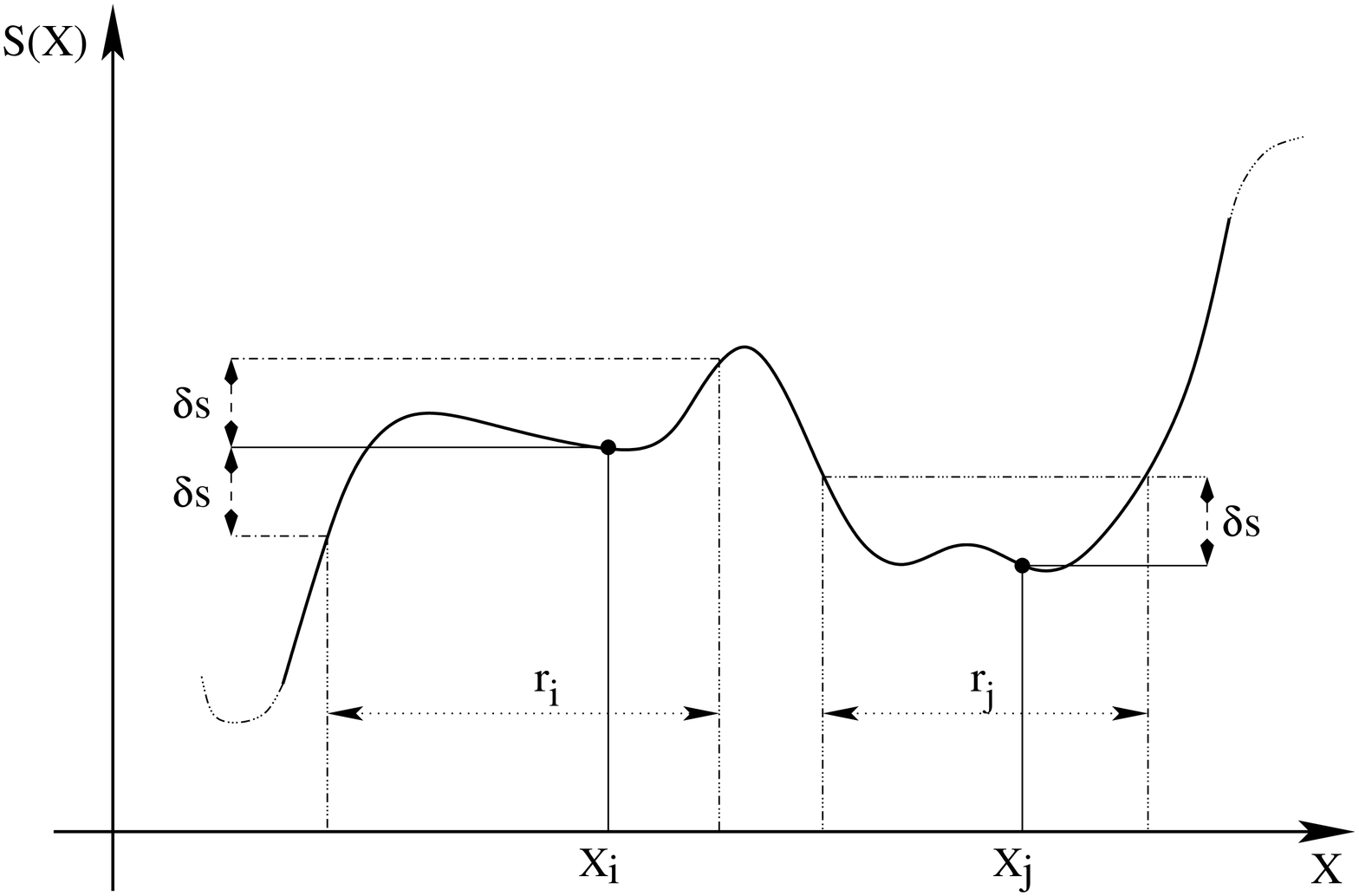,width=0.47\textwidth,height=0.34\textwidth}}
\vspace{0.2cm}
\caption{A pictorial representation of the exit-distance method.
$X_i$ and $X_j$ are two points picked at random and $r_i$ and $r_j$
are the corresponding exit distance from the barrier $\delta s$.}
\end{figure}
We fix a value for the
signal fluctuation, $\delta s$, then we pick at random a reference
point $x_0$ and measure the first forward ($|s(x_0 + r_f)-s(x_{0})| \ge
\delta s$) and backward ($|s(x_0 - r_b)-s(x_{0})| \ge \delta s$) exit from the
barrier, $r(\delta s)$. Then we put $r(x_0,\delta s) = r_b+r_f$.
See Fig.~2 for a pictorial view of the method.
Repeating the observations for many point $x_0$ and for different
barrier heights, we can define the inverse structure functions \cite{Je99} as 
\be
T^{(p)}(\delta s)= \langle r^p(\delta s) \rangle \sim \delta
s^{\chi(p)}\,,
\label{eq:invss} 
\ee
where the average is taken with respect to the random choice of the
point $x_0$ \cite{nota}.
For the case of simple signals such as (\ref{smooth}), a rigorous
estimate of the scaling exponents of inverse statistics moments can be
derived as follows. If the signal spectrum is $E(k)\sim k^{-\alpha}$
with $3 \le \alpha\,<\,5$, we can write the signal increment as
\begin{equation}
   s(x+r)-s(x) \sim \frac{ds(x)}{dx}\,r + c(x)\,r^h \,.
   \label{local}
\end{equation}
Here we have only kept the two most important scaling behaviors:
$O(r)$ because of the differentiability and $O(r^h)$ from the spectrum
exponent. The scaling exponent $1 \le h < 2$ is related to the
spectrum slope by the dimensional relation $\alpha=1+2h$, while $c(x)$
is a continuous function of $x$. By studying the exit event, in the
limit of a small barrier height, we may observe two different kinds of
event. The first, with probability one, is the differentiable scaling
$r(\delta s ) \sim \delta s$. The second, observed at those points
where the first derivative vanishes, is the subleading behavior,
$O(r^h)$, in (\ref{local}). \\ One may estimate the
probability of this second situation as follows.  With $3 \le \alpha < 5$
the first derivative is a self-affine signal with H\"older exponent $
\xi = (h-1) <1$, which vanishes on a fractal set of dimension $D =
1-\xi = 2-h$.  Therefore, the probability to see the sub-leading term
$O(r^h)$ dominating the exit event in (\ref{local}) is given by the
probability to pick at random a point on a fractal set of dimension
$D$, i.e.,
\begin{equation}
P( r \sim (\delta s)^{1/h} ) \sim r^{1-D} = (\delta s)^{1-1/h}.
\end{equation}
Taking into account both events, we end with the following 
{\it bifractal} prediction for inverse statistics moments\,:
\begin{equation}
T^{(p)}(\delta s) \sim (\delta s)^{\chi_{\rm bf}(p)},\; \chi_{\rm bf}(p) = 
\min\!\left(p,\frac{p}{h}+1-\frac{1}{h}\right).
\label{bifrattal}
\end{equation}
From (\ref{bifrattal}), we conclude that laminar differentiable
fluctuations influence the inverse statistics only up to moments of
order $p=1$; for larger $p$, the PDF is dominated by the sub-dominant
behavior, $(s(x+r)-s(x))\sim r^h$.  In other words, the extrema of the
signal play the role of singularities for the inverse statistics\,:
close to the extrema, events with much longer exit distances are
observed when $\delta s \rightarrow 0$. For one-dimensional signals as
(\ref{smooth}) the prediction is verified with high accuracy
(see~Fig.~3).\\ In the general multiaffine case, signal increments
scale as $\delta_r s(x) \sim r^{h(x)}$ with probability $P_r(h) \sim
r^{\,1-D(h)}$, where the function $D(h)$ can be interpreted as the
fractal dimension of the set where the H\"older exponent $h$ is
observed \cite{PF85}.\\ For such a signal, it is possible to
obtain~\cite{Je99,BCVV99} a link between the inverse statistics
exponents, $\chi(p)$, and the fractal dimension, $D(h)$:
\begin{equation}
   \chi(p) = \min_h \left({{p + 1 - D(h)} \over h} \right)\;.
   \label{legendre2}
\end{equation}
In the case of the smooth signal (\ref{smooth}), one can see that
(\ref{legendre2}) coincides with the bifractal prediction
(\ref{bifrattal}), as soon as we write $D(h) = 2-h$ for $h
\equiv (1,(\alpha-1)/2)$.
\begin{figure}[hbt]
\label{fig:inverse1d}
\centerline{\epsfig{figure=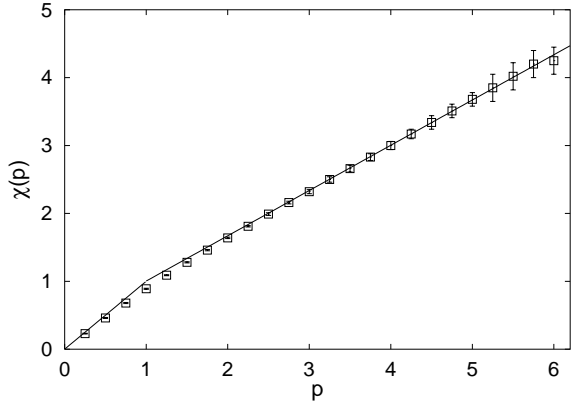,width=0.45\textwidth,angle=0}}
\protect\caption{Scaling exponents $\chi(p)$ for the $1D$ signal (\ref{smooth}) with $\alpha=4$. The solid line
gives the bifractal behavior. Moments have been
computed using $10^3$ realizations of the signal (\ref{smooth})
with $2^{17}$ modes; for each realization
$2^{12}$ starting points, $x_0$, have been taken at random.}
\end{figure}
\subsection{Second Difference structure functions}
\label{sec:2.2}
Another way to eliminate the trivial differential scaling and
extract some statistical information from smooth signals has been
suggested in \cite{Eyink}. The basic idea is to consider
moments of the second difference $\Delta_r s \equiv
(s(x+r)+s(x-r)-2\,s(x))$, so that the differentiable contribution,
$\delta_r s \propto r$, is automatically eliminated.  For the signals
under investigation, we have that at the leading order $\Delta_r s \sim r^h$ with $1 \le h < 2$, and moments behave as \be S^{(p)}_{SD}(r)\equiv \langle |\Delta_r s|^p \rangle \sim
r^{z_p}.  \label{def:sfeyink} \ee 
In the monofractal case (globally
self-similar signals), one expects $z_p=ph$. The analysis done for the 
same stochastic $1D$ signal of
 (\ref{smooth}) with
$h=1.5$, confirms this expectation (see Fig.~4).\\
In the general case, i.e., when many more-than-differentiable
fluctuations are present, the scaling exponents $z_p$ are non-trivially related to the distribution of the $h$ exponents.  The
difficulty to give a multifractal prediction for $\Delta_r s$
increments stems from the fact that it is a three-point quantity,
depending on the simultaneous fluctuations between ($x$, $x-r$) and
($x$, $x+r$). Therefore, to draw the multifractal picture, we would need in
addition a complete control of the spatial correlations. Something
which is in principle feasible \cite{ONM93,BBT97} but which goes
beyond the aim of this paper.
\begin{figure}[hbt]
\label{fig:eyink1d}
\centerline{\epsfig{figure=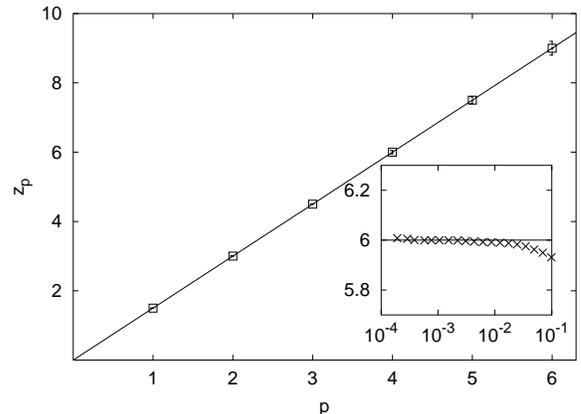,width=0.45\textwidth,angle=0}}
\protect\caption{Scaling exponent for the $1D$ signal (\ref{smooth}) with
the same parameters as in Fig.~3, the straight
line shows the expected behavior $z_p=hp$ with $h=1.5$. The inset
shows the local slope for $p=4$.}
\end{figure}
\section{Spatial statistics in $2D$ smooth velocity fields}
\label{sec:3}
Let us now analyse the inverse and second difference statistics of the
two dimensional velocity field obtained by DNS of the Navier-Stokes equation (\ref{eq:NS}). We performed four different
sets of numerical experiments, with periodic boundary conditions on a
spatial grid of $1024^2$ collocation points. In all of them, we
considered a Gaussian forcing, $\delta$-correlated in time, with support
in a restricted band of wave numbers $4 < k_f \leq 6$.
\begin{figure}
\label{fig:spettro}
\centerline{\epsfig{figure=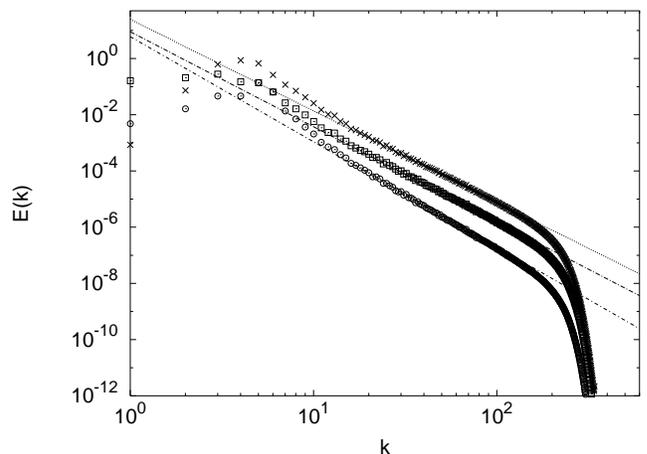,width=0.5\textwidth,angle=0}}
\protect\caption{Log-log plot of the velocity spectra for 
two different drag coefficients with the Eckman linear friction, 
run B (middle) and run  C (bottom), and with a 
hypodiffusive friction, run  D (top). Straight-lines correspond to
the best fit power laws,  $k^{-3.38}$, $k^{-3.74}$ and $k^{-3.26}$ respectively. Run A is not shown because it is almost 
indistinguishable from run D.}
\end{figure}
In three out of four simulations,  
we used Eckman linear friction, i.e. $\rho=0$ in (\ref{eq:NS}) with different coefficients (simulations A,B,C, in the following). In the
fourth run, we used a hypo-diffusive term at large scales, 
$\rho=2$, (referred as case D in what follows). 
Table 1 is a summary of the DNS parameters, 
together with the best-fit spectrum exponent $\alpha$
for all runs.\\
In Fig.~5 we show the averaged velocity spectrum for run B,C and D (run
A gives a slope almost coincident with that of run D).
By comparing them, it is evident 
that the spectrum slope depends
on both the drag coefficient (runs A,B and C) and 
on the drag mechanism (run D).\\ Evidently, 
we are in presence of large-scales effects which 
somehow affects small scales velocity fluctuations.  Let us try to 
quantify this statement by using the  inverse statistics analysis. 
First we compare the inverse structure functions measured on several
snapshots of the DNS, with those obtained after randomization of all
velocity phases on the same frames. 
The rationale for this test is to investigate the
importance of correlations between fluctuations at different
wave-numbers and therefore the ``information'' content brought by
coherent structures in 2D turbulent flows.\\ 
If we look at a one-dimensional cut of the velocity field, 
before and after phases randomization, it is rather difficult to
distinguish the original DNS field from that-one with randomized phases. 
This is due to the steepness of the spectrum, i.e., only few modes dominate the
real-space configuration. Despite the apparent similarity big
 differences arise when looking at inverse moments.\\
Because of the limited numerical resolution, the only 
quantitative statements one can give are for relative scaling properties. 
Therefore, we measure scaling laws of the inverse
statistics by plotting all moments $T^{(p)}(\delta u)$ versus a
reference one, say $T^{(2)}(\delta u)$.  This is the same technique
called ESS~\cite{ess}, fruitfully applied in the analysis of
$3D$ turbulent data with the aim of re-absorbing some finite size
effects and extracting scaling information also at moderate
resolution. Therefore, we concentrate on the following
relative scaling properties\,:
$$
T^{(p)}(\delta u) \propto (\,T^{(2)}(\delta u)\,)^{\chi(p)/\chi(2)}\,.
$$ 
In Figs.~6 we summarize our findings. 
Inverse moment exponents, $\chi(p)/\chi(2)$, measured on the
turbulent fields with randomized phases follow the bifractal
prediction (\ref{bifrattal}), with the value of $h$ extracted from the
spectrum (see Table 1). Conversely, the longitudinal and transversal
inverse-statistics moments without phases randomization show anomalous
 scaling laws,  which deviate from the bifractal law given in 
(\ref{bifrattal}). 
In Figs.~6, we show the curve
$\chi(p)/\chi(2)$ for both randomized and non-randomized transversal
exit moments for run C and D.  For $p<1$, the statistics of the
randomized data and that of the turbulent data almost coincide being
those moments (with $0< p<1$) dominated by the laminar fluctuations
$u(x+r)-u(x) \sim r$.  To better appreciate differences in the scaling
curves, we show in the inset of Figs.~6 the local slopes of
$T^{(4)}(\delta u)$ versus $T^{(2)}(\delta u)$, for the randomized and
non randomized data.\\ The following scenario can be drawn. Randomized
data follow the bifractal prediction, while the non randomized ones
are definitely different and display anomalous scaling. Moreover, the
anomalous scaling is present for all choices of the drag mechanism.\\ 
\begin{figure}
\label{fig:inverse-long}
\centerline{\epsfig{figure=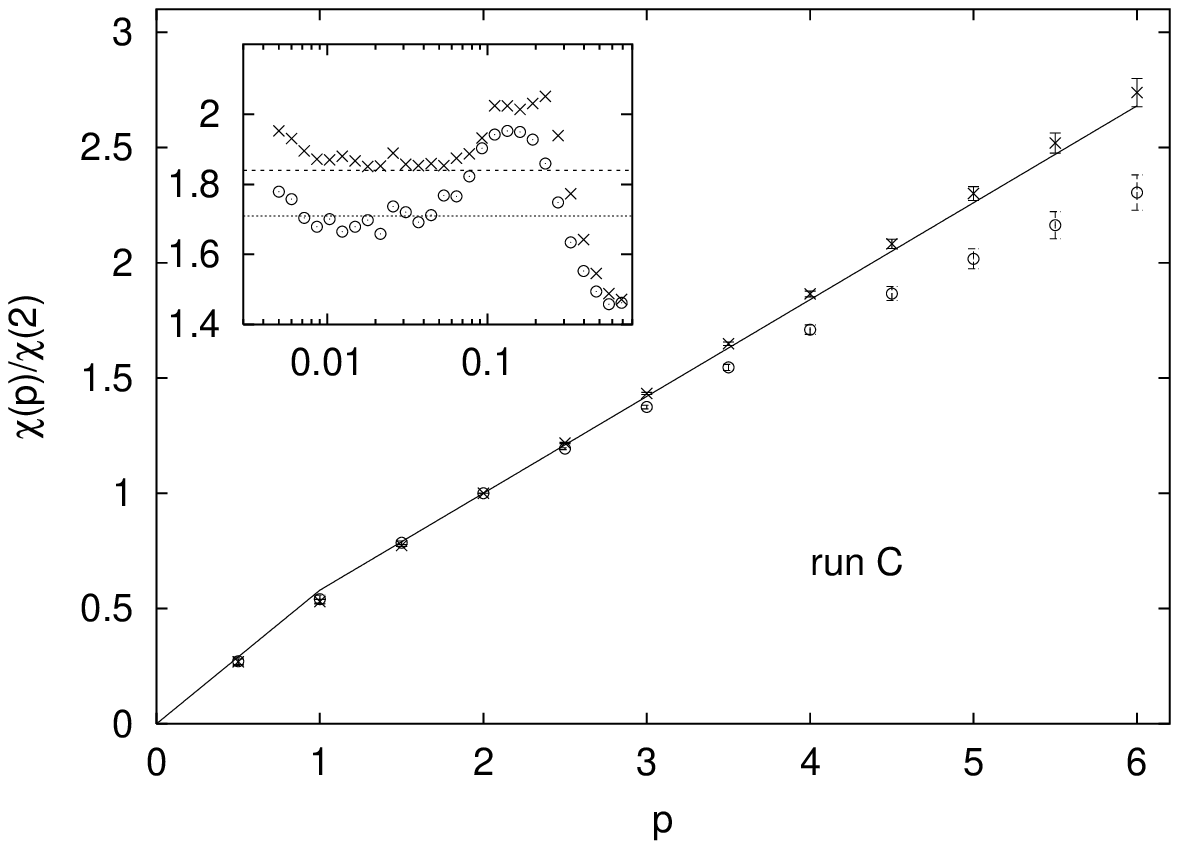,width=0.45\textwidth,angle=0}}
\centerline{\epsfig{figure=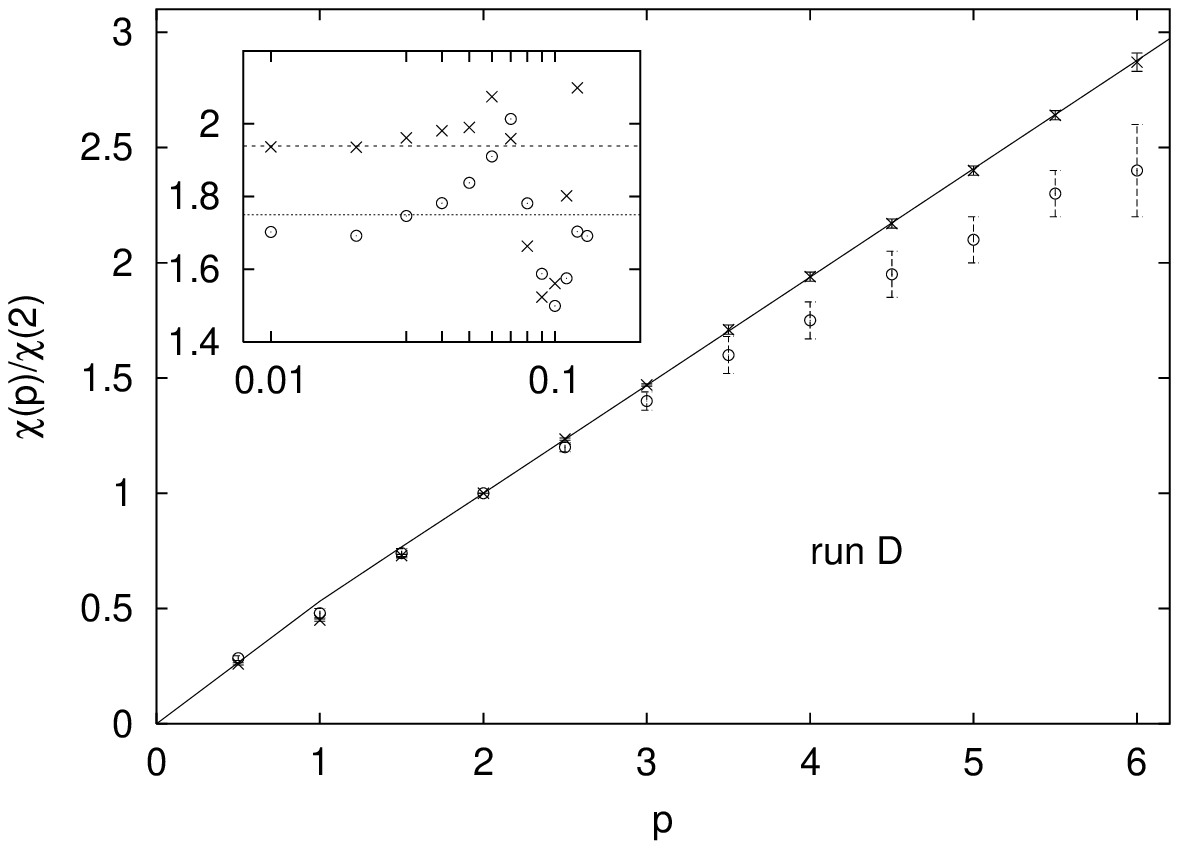,width=0.45\textwidth,angle=0}}
\protect\caption{$\chi (p)/\chi (2)$ for exit moments obtained in run
C (top) and D (bottom) ($\circ$), compared with the same data after
phases randomization ($\times$). In both figures, the solid line
corresponds to the bifractal prediction.  The values of $h$ used for
the theoretical prediction are obtained from the spectrum scaling
exponents (see Table 1), namely $h=1.38$ (run C) and $h=1.13$ (run D).
Errors on scaling exponents have been computed according to
the fluctuations of the local slopes. In each inset, it is shown the
local slope of $T^{(4)}(\delta u)$ versus $T^{(2)}(\delta u)$ for the
true signal ($\circ$) and the randomized one ($\times$).}
\end{figure}
For Second Difference statistics analogous results have been
found, that is a monofractal behavior for the randomized field and an
anomalous behavior for the turbulent one. In Figs.~7 we show the
scaling exponents $z_p$ for run (C) and run (D). Longitudinal and
transversal components, within the errors, coincide. The SD analysis
confirms that the statistics of laminar events for the $2D$ turbulent
velocity field displays a complex, multifractal structure. \\
Concerning the case of run C, i.e., with linear friction, it is
interesting to compare the results of the Second Difference moments
with some recent analytical results \cite{bernard}. In \cite{bernard},
it is argued that in presence of linear Eckman friction, the second
and third order (standard) structure functions behave as
$S_2(r)=\langle \delta u ^2(r) \rangle \sim a r^2+b
r^{2+(\alpha-3)}$ and $S_3(r)=\langle (\delta u (r))^3 \rangle \sim d
r^3+e r^{\alpha}$, being $\alpha>3$ the spectrum slope, and $a,b,d,e$
some constants. From these results, it is easy to extract the exponents of
the SD moments, i.e., $z_2=\alpha-1$ and $z_3=\alpha$. Actually our
data give slightly larger values. Such discrepancies may be due to
strong finite Reynolds effects which, in $2D$, are particularly severe
due to the interplay between inverse cascade and friction in the low
$k$ region of the spectrum.\\
\begin{figure}
\label{fig:eyink}
\centerline{\epsfig{figure=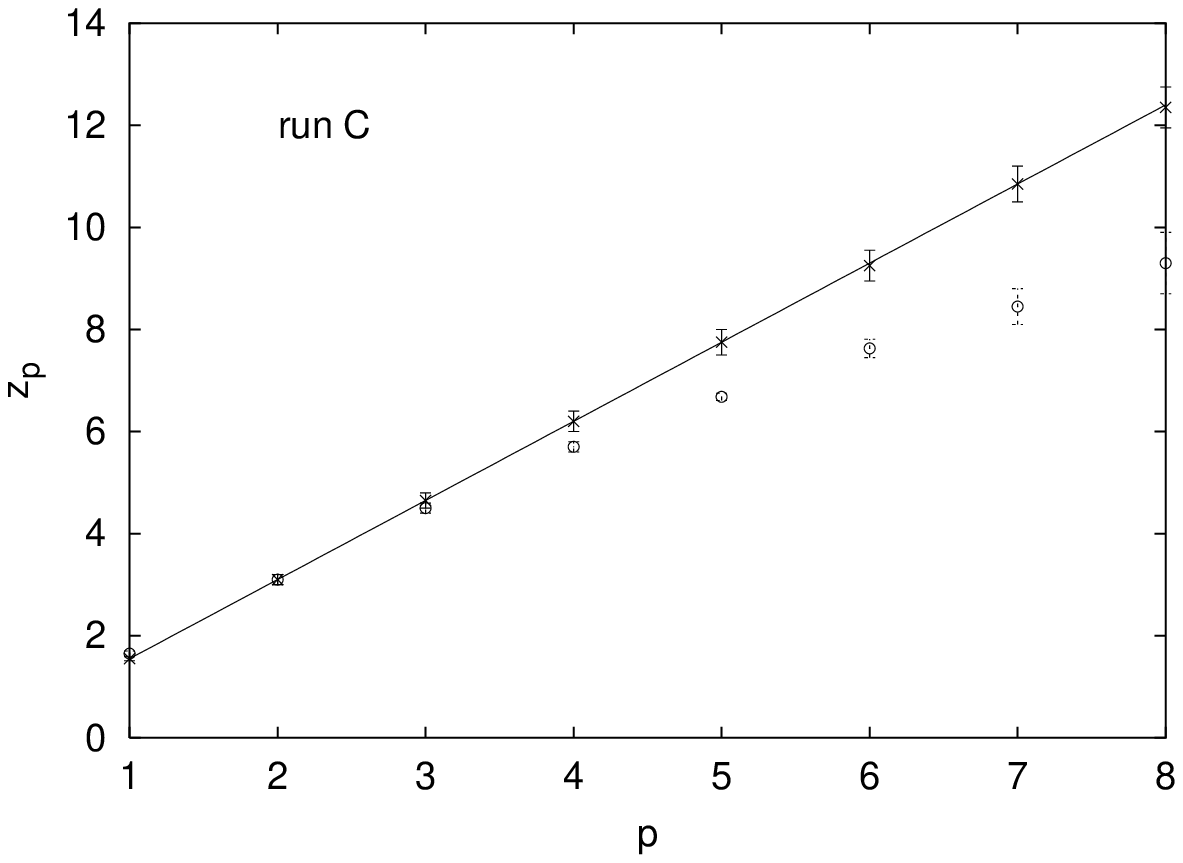,width=0.45\textwidth,angle=0}}
\centerline{\epsfig{figure=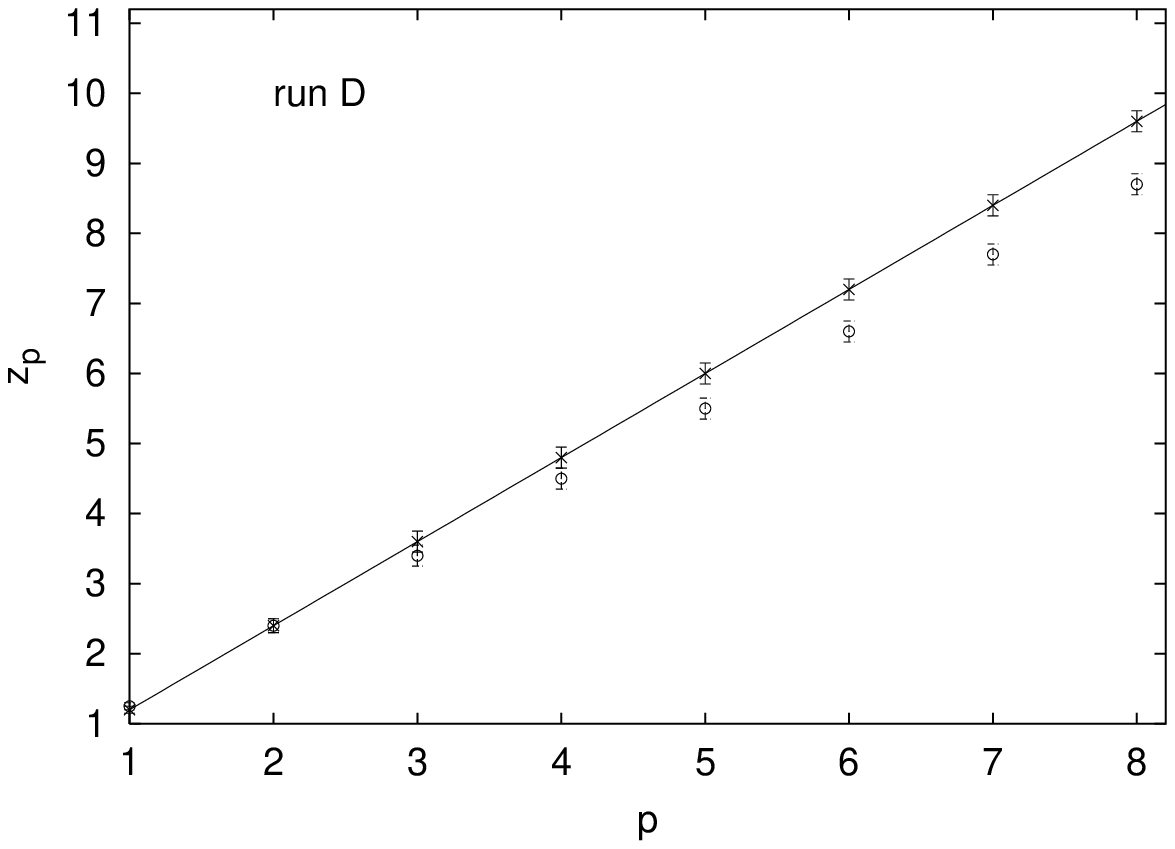,width=0.45\textwidth,angle=0}}
\protect\caption{(top) Second Difference exponents, $z_p$, for 
moments obtained in run (C)
($\circ$) and after randomization ($\times$); (bottom) the same but
for run D. The straight lines correspond to the monofractal behaviour
$z_p=\tilde{h}p$ with $\tilde{h}=1.55$ (run C) and $\tilde{h}=1.20$
(run D).}
\end{figure}
\section{Temporal statistics}
\label{sec:4} 
As it is well known, in $3D$ turbulence we can recast the temporal
behavior of the velocity field into the spatial domain via the Taylor
hypothesis (frozen turbulence hypothesis): the effect of large scales
is just that of a uniform sweeping which does not modify the small
scale structures and their energy content. In $2D$ the absence of a
time hierarchy rules out such a possibility
\cite{kraichnan_2d,Bo93}. This is also evident by looking at snapshots
from numerical simulations, which show that the time evolution of the
dynamics is dominated by stable, long-lived structures (see Fig.~1).
For such a reason it is non trivial to predict the velocity temporal
statistics collected in a fixed spatial location.\\ We performed a DNS 
of (\ref{eq:NS}) taking as a large-scale forcing
a function $F$ of constant amplitude at some characteristic
wave-numbers  $4 <|k_f| \leq 6$ and time-independent. We performed a long time integration of the $2D$ N-S equations,
at resolution $512^2$ and collected statistics for hundreds large
eddy turn over times, estimated as $t_{eddy} \approx 1/ \om_{rms}$
(details on the numerical simulation can be found in \cite{notaSP}).
Once the system reached a stationary state, we started to collect the
time evolution of the velocity fields at some specific spatial
locations with a sampling time $\tau_{samp} \sim 2.5\cdot 10^{-2}
t_{eddy}$. 
\begin{figure}
\label{fig:segnali}
\centerline{\epsfig{figure=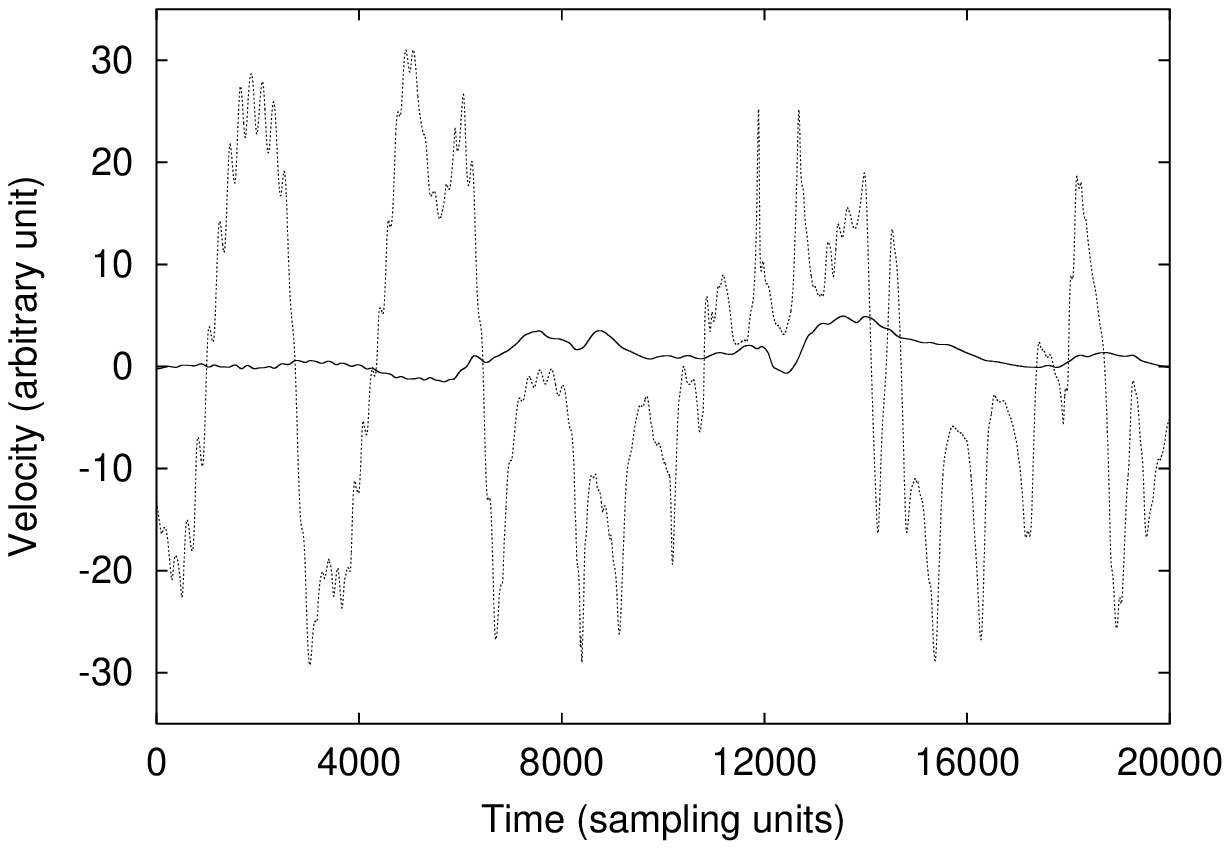,width=0.45\textwidth,angle=0}}
\centerline{\epsfig{figure=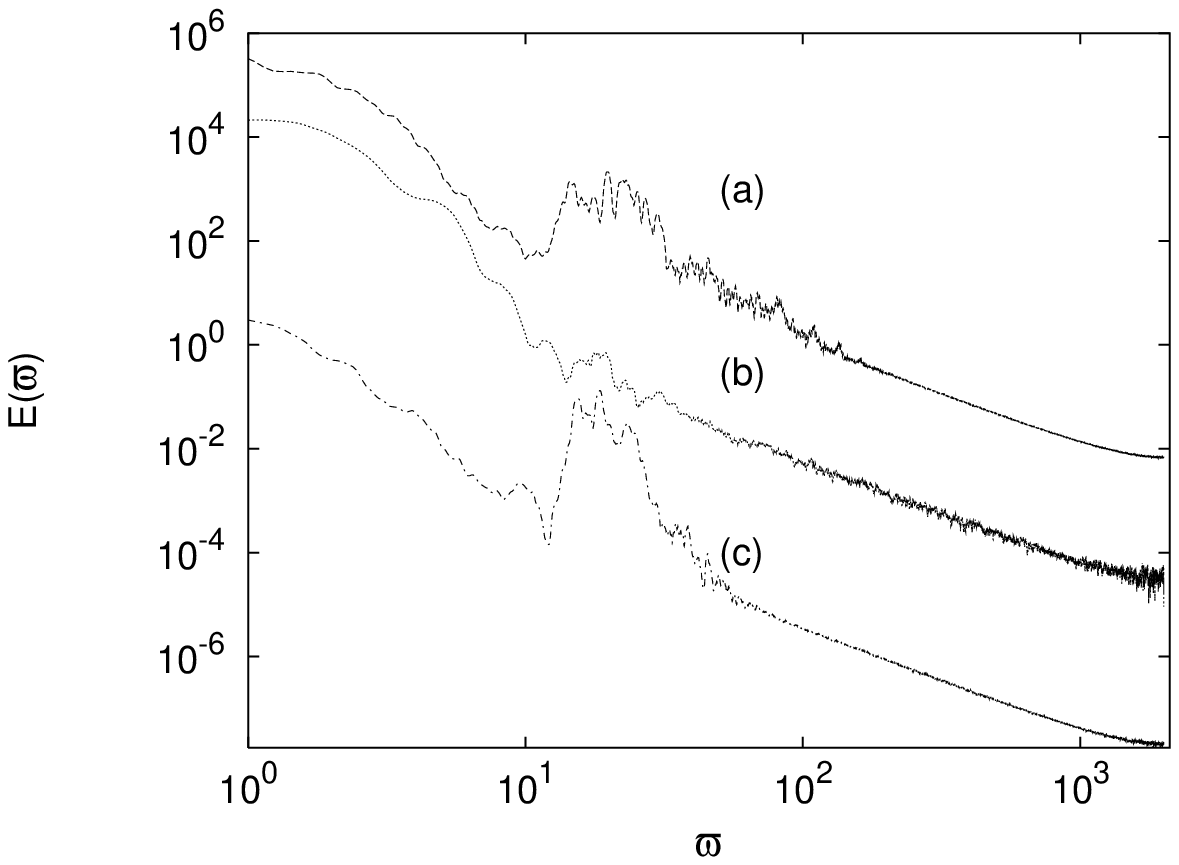,width=0.45\textwidth,height=0.225\textheight,angle=0}}
\protect\caption{(top) Time recording of the x-component of the
velocity field of the $p_{in}$ probe (dotted line) and the $p_{out}$
probe (solid line);(bottom) log-log plot of the frequency spectra of the following signals: 
(a) $E_{p_{in}}(\varpi)$ of the probe near a
coherent structure; (b) $E_{p_{out}}(\varpi)$ of the laminar one;
(c) $E_{k_f}(\varpi)$ of the time evolution at a particular Fourier mode $\k$, in 
forced wave-number band $|k_f|=(4;6]$. }
\end{figure}
Some observations are noteworthy. The first one
concerns the {\it ergodicity} of the velocity field ${\bm u}({\bm
x},t)$. Temporal signals collected at different spatial locations
possess different probability distribution functions. In particular
the range of variations of the local {\it rms} values $u_{rms}({\bm
x_0},t)$ is so wide that we can not average time histories recorded at
different points. It is difficult to say if waiting long enough one
would recover, as expected, some stable ergodic properties. Certainly,
with our statistics we feel confident to report results only on local
averages, avoiding to mix temporal evolutions in different
spatial locations.\\
In particular, we chose to report those describing two typical
spatial situations: one, $p_{in}$, situated in the core of a vortical
structure, the other, $p_{out}$, in a laminar region. This means that
notwithstanding the turbulent evolution of the field, the motion of
the vortices is so slow that probes almost maintain their respective
``character'' ({\it in} and {\it out} of a vortex) all the simulation
long.\\ Looking at a sample of the time series recorded by the two
probes, the signals seem very different and change when the probe
passes from a laminar region to a vortical one (see Fig.~8 (top)). To
have a better understanding, it is useful to consider the frequency
spectra $E(\varpi)$ of the signals, calculated from the temporal
Fourier transform of the stationary time correlation function, e.g.,
of $C(u_x(p),\tau)= \la \,u_x(p,t+ \tau) \,u_x(p,t)\,\ra$.\\ In Fig.~8
(bottom), we can observe the spectra of the two probes. At
variance from the spatial spectra, it is not possible to extract a
clear scaling behavior. One can only identifies an exponential decay,
and a peak region located at the frequency $\varpi^{(L)} \approx 0.5
1/t_{eddy}$ (here and in the sequel ``$(L)$'' stands for ``large
scale''.)  It is easy to recognize that $T^{(L)}$, defined as
$T^{(L)}\equiv 1/\varpi^{(L)}$, is the typical time scale associated
to the large-scale structures, either estimating it from the vorticity
content of the largest structures $T^{(L)}=1/\sqrt{\la \om^2 \ra}$ or
from their typical revolution time. In other words, Fig.~8 (bottom)
tell us that in each spatial point the time evolution is governed by
the typical oscillation frequency of the forced large-scale
structures.\\ This is confirmed by the comparison of the spectra
probes with the spectra built from the time correlation of the Fourier
transformed velocity field $\hat{\u}(\k,t)$, at a given mode, $\k_f$,
belonging to the forced wavenumber band. Indeed, all
spectra posses a peak at  frequency $\varpi^{(L)}$.\\ We now pass
to investigate direct and inverse statistics of $\u(t)$. Direct structure
functions behave trivially for both probes, $ S_p(\tau)= \la [u({\bf
x_0},t+\tau)- u ({\bf x_0},t)]^p \sim c_p\,\tau^{p},\,$ where $u$ can
be either one of the components $(u_x,u_y)$ or the velocity
modulus.\\
It turns out that inverse temporal statistics does not posses good
scaling laws. Therefore, we refrain from giving any quantitative
statement while we concentrate on some qualitative properties showed
by PDF's of temporal inverse events measured at the two probes,
$p_{in}$ and $p_{out}$.\\ In Fig.~9 we plot, for the probe $p_{in}$,
various PDFs $P \left(\tau_{\delta u}/\la\tau(\delta u)\ra\right)$, at
varying $\delta u$, all re-scaled with their mean value\ $\la
\tau(\delta u) \ra$. First, we notice that PDFs collapse 
very well, indicating the absence of intermittency
effects. Second, between the peak and the exponential tails at large
$\tau$, each probability density function exhibits, on a wide range of
scales, a power law behavior $P(\tau_{\delta u}) \sim (\tau_{\delta
u})^{-\gamma}$ with an estimated exponent $\gamma \approx -1$. 
On the other hand, PDFs measured on the probe outside the 
vortex, $p_{out}$, show a very different qualitative trend (inset
of Fig.~9).
In particular, there is not any clear power law behavior. This indicates 
that very large exit events become less and less probable 
outside the probe, something which must have 
to do with the absence of very smooth fluctuations in the
vortex background.  
\begin{figure}
\label{fig:pdfsinv041}
\centerline{\epsfig{figure=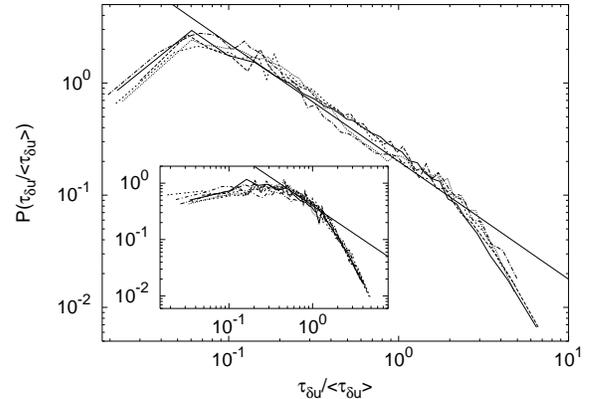,width=0.45\textwidth,angle=0}}
\protect\caption{Exit-time probability density functions $P(\tau_{\delta u})$ 
 measured on the statistics of the probe $p_{in}$. The curves, calculated for 
different exit barriers $\delta u$, are normalized by their first moment.
In the inset, the same for the probe $p_{out}$.
In each figure, the straight line is the power law behavior $\tau_{\delta u}^{-1}$.}
\end{figure}
 Altough qualitative, the inverse-statistics properties 
 allow to distinguish among different temporal 
statistical behaviors associated to different fluid regions.
\section{Conclusion} 
\label{sec:5} 
To summarize, we have studied inverse statistics moments for signals
with a more than smooth spectrum, i.e., signals which are
differentiable and with non-trivial stochastic sub-leading
fluctuations. We have shown that statistical velocity properties of 2D
turbulent flows are not simply described in terms of the spectrum
slope. From the exit-distance analysis it is possible to highlight a
whole spectrum of more-than-differentiable fluctuations. These, being
connected with laminar events, are the strongest statistical signature
of the large-scale coherence. Experiments with different methods of
removing/pumping energy at large scales should be performed, to
investigate the importance of large-scale structures in the inverse
statistics of flows with different spectra.  We have quantified
laminar fluctuations also by using Second Differences, i.e., direct
velocity increments subtracted of their linear differentiable
behavior. We have found also in this case that
more-than-differentiable fluctuations are not simply described by one
single exponent.\\ As a final remark, we stress that inverse
statistics provide a completely new statistical indicator with respect
to the standard direct statistics observables. We have shown that such
method is necessary in all those cases where non-trivial fluctuations
are sub-leading with respect to the differentiable
contributions. Obviously, the same kind of analysis here reported can
be extended to other temporal signals, applying the method to a broad 
class of natural phenomena. As an example, we just mention possible 
applications in situations common to
climatology or meteorology where estimating the probability of
persistent velocity configurations, or of any other dynamical variable,
is relevant. As a perspective, an important generalization is the
investigation of multi-dimensional signals by studying the statistics
of $d$-dimensional volumes between equispaced iso-surfaces.
\section{Acknowledgments}
We thank A.~Vulpiani for his contribution in the early stage of this
 work. We acknowledge useful discussions with R.~Benzi, G.~Boffetta and 
G.~Eyink. This work has been partially supported by the EU under 
the Grant No. HPRN-CT 2000-00162 ``Non Ideal Turbulence'' and the Grant
 ERB FMR XCT 98-0175 ``Intermittency in Turbulent Systems''. 
M.C. is partially supported by the European Network 
"Non Ideal Turbulence" (contract number HPRN-CT-2000-00162). We also 
acknowledge INFM  support (Iniziativa di Calcolo Parallelo).

\begin{table}[htb]
\narrowtext
\caption{Drag parameters $\rho, \beta_{\rho}$, spectrum
slope $\alpha$,  and the real space sub-leading 
scaling exponent, $h = (\alpha -1)/2$
 for the various numerical experiments.
 The value of each $\alpha$ has been
obtained by a best fit in the region $|\k| \approx (20-100)$ (see
Fig.~5).  By performing the fit  in
the region $|\k| \approx (15-60)$, slightly larger values of $\alpha$
are obtained.  These discrepancies can be associated to the interplay
between the inverse cascade of energy and the friction acting on it,
which contanimates the upper part of the spectrum.}
\begin{tabular}{ccccc} 
DNS label & $\rho$ & $\beta_{\rho}$ & $\alpha$  &  $h$ 
 \\ 
\hline 
A & 0 & 0.01 & 3.26(6) & 1.13 \\ 
B & 0 & 0.10 & 3.38(8) & 1.19 \\ 
C & 0 & 0.30 & 3.74(8) & 1.37 \\ 
D & 2 & 14.0 & 3.26(6) & 1.13
\label{table:1}
\end{tabular}
\end{table}

\end{multicols}


\begin{thebibliography}{99}
\bibitem{nota} It can be shown that moments calculated in
this way are the same as those calculated from a sequential average, i.e.
``scanning'' consecutively the time series and then using a suitable
renormalization, see \cite{BCVV99} for details.
\bibitem{notaSP} For sake of completeness, we recall that in this
series of numerical simulations, enstrophy is dissipated by an
hyper-viscosity term with $q=4$, while energy is removed using an IR
drag term with $\rho=2.$ The resulting energy spectrum has the form
$E(k) \sim k^{-\alpha}$, with $\alpha=5.0 \pm 0.2$.
\end{thebibliography}
\end{document}